\documentclass[a4paper,12pt]{article}
\usepackage{amsmath,amssymb,amsfonts}
\usepackage[utf8]{inputenc}
\usepackage[english,russian]{babel}
\usepackage[]{fontenc}
\usepackage{pst-circ}
\usepackage{mathrsfs}
\usepackage{breqn}
\usepackage{bm}
\usepackage{pb-diagram}
\def\A{\mathcal A}
\def\AA{\tilde{A}}
\def\Alt{\operatorname{A}}
\def\around{\circlearrowleft}

\def\beq{\begin{equation}}
\def\eeq{\end{equation}}
\def\beqq{\begin{equation*}}
\def\eeqq{\end{equation*}}

\def\Cs{{\C^s}}
\def\Css{{\C^s}^*}

\def\D{\mathcal D}
\def\der{\partial}
\def\dder{d}
\def\ds{\displaystyle}

\def\E{\mathcal{E}} 
\def\End{\operatorname{End}}
\def\F{{\bm F}}
\def\Fo{\mathcal{F}^s}
\def\Fstar{\mathcal{F}^{s*}}
\def\fnormal{\vdots}
\def\T{\mathcal{T}} 

\def\gl{\mathfrak{gl}}

\newcommand\Heis[1]{\H^{#1}}
\def\H{\mathcal{H}}
\def\HH{\H^s_-}
\def\Hz{\HH(z)}
\renewcommand{\L}[1]{\Lambda^{s,{#1}}}
\def\Ls{\hat{\Lambda}^{(s)}}

\newcommand{\pk}[1]{p_{#1,k}}
\def\ppi{\bar{\pi}}

\newcommand\Q[1]{Q_{#1}}
\newcommand\QQ[1]{Q^{-1}_{#1}}
\newcommand\q[1]{q_{#1}}
\def\Qoper{\hat{Q}}

\def\S{\mathcal{S}}

\renewcommand\U[1]{\Psi_{#1}}
\def\UUU{{\bm \Psi}^*}
\newcommand{\UUUU}[1]{{\bm \Psi}^{*(#1)}}
\def\UU{{{\bm \Psi}}}

\newcommand\V[1]{\Phi_{#1}}
\def\VVV{{\bm \Phi}^*}
\newcommand{\VVVV}[1]{{\bm \Phi}^{*(#1)}}
\def\VV{{{\bm \Phi}}}
\def\vf{\varphi}

\def\Y{\operatorname{Y}}

\def\C{{\mathbb{C}}}
\def\N{{\mathbb{N}}}
\def\Z{{\mathbb{Z}}}	

\newcommand{\rf}[1]{(\ref{#1})}

\newtheorem{lemma}{Lemma}[section]
\newtheorem{proposition}{Proposition}[section]

\begin{document}
	\bigskip
	\hfill{ITEP-TH-28/19}
	\begin{center}
		{\Large\bf Matrix elements of vertex operators and fermionic limit of Spin Calogero-Sutherland system}
		
		\bigskip
		{\bf S.M. Khoroshkin$^{\circ\star}$,  \, M.G. Matushko$^{\circ\ast}$
		}\medskip\\
		$^\circ${\it National Research University Higher School of Economics, Moscow, Russia;}\smallskip\\
		$^\star${\it Institute for Theoretical and Experimental Physics, Moscow, Russia;}\smallskip\\ 
		$^\ast${\it  Center for Advanced Studies, Skoltech, Moscow, Russia}
	\end{center}
	\selectlanguage{english}
\begin{abstract}\noindent
We present a construction of an integrable model as a projective type limit of spin
Calogero-Sutherland model with $N$ fermionic particles, where $N$ tends to infinity. It is implemented in the multicomponent fermionic Fock space. Explicit formulas for limits of Dunkl operators and the Yangian
generators are presented by means of fermionic fields.
\end{abstract}
	\section{Introduction}
 Recent researches on limits of  Calogero-Sutherland (CS) systems \cite{NazSk1,NazSk,SerVes} followed the cornerstones of the theory of symmetric functions, where  the ring $\Lambda$ of symmetric functions on the one side is regarded as the projective limit of the spaces of symmetric polynomials in finite number of variables \cite{Mac}, 
 and on the other side is a representation of the Heisenberg algebra $\H$ and thus can be related to the Fock space. Operators, forming the Hamiltonians of the finite 	CS system can be  rewritten in a way admitting a translation in the Fock space. 
 
  These constructions heavily  use an equivariant family of Heckman-Dunkl \cite{Heck} operators as a counterpart of the Lax operator in integrable systems. 
  The procedure requires a creation of an auxiliary space of functions symmetric over all variables except one \cite{NazSk1,NazSk} or as a variant of polynomials in one variable with coefficients being symmetric functions of all variables \cite{SerVes, N}. These spaces are created by the vertex operator
  $$\Phi_+(z)=\exp\sum_{n\ge0}z^n\frac{\partial}{\partial p_n}$$
  which use in the finite system is nothing more than the Taylor expansion. Here $p_n=\sum_kx_k^n$ are Newton polynomials.
  Instead of taking trace of the Lax operator here one sums up the images of all powers of Dunkl operators. The latter is equivalent to the integration with the weight function $\varphi_-(z)$ being the negative part of the bosonic field
  $$ \varphi_-(z)=
  \sum_{n>0}\frac{p_n}{z^n}.$$
  
  The spin CS system has more symmetries, which constitute the representation of the Yangian $\Y(\gl_s)$.
  The integrals of motion of the system can be chosen as coefficients of the related quantum determinant. We provide a description of finite CS system in Section 2.   
	This limiting bosonic construction system was generalized in \cite{KMS} to spin CS system using the language of polysymmetric functions, that is polynomials, symmetric on the groups of variables. See also \cite{N}. 
	
	In \cite{KM} analogous ideas were applied to CS system of fermionic particles. The finite object is the same  -- the ring $\Lambda$ of symmetric functions,
	but its finite reincarnations are different: these are antisymmetric polynomials. Then the Vandermonde $\prod_{i<j}(x_i-x_j)$ brings the appearance of the full vertex operator 
	$$
	\Psi(z)=z^{p_0}\exp\left(-\sum_{n>0}\frac{p_n}{n z^n}\right)\exp\left(\sum_{n\ge0}z^n\frac{\partial}{\partial p_n}\right)
	$$ 
	in the creation of the auxiliary space. The weight function of the integration is now
	$$
	\Psi^*(z)=z^{-p_0}\exp\left(\sum_{n>0}\frac{p_n}{n z^n}\right)\exp\left(\sum_{n\ge0}-z^n
	\frac{\partial}{\partial p_n}\right).
		$$

	The fermionic limit of spin CS system was studied by D. Uglov. In his fundamental research  \cite{Uglov} he constructed the integrable system being an inductive limit of finite-dimensional spin CS system.	
	In this paper we suggest and develop a different approach, realized as a projective type limit. It leads to the limiting system which differs from \cite{Uglov}. We start from the fermionic Fock space $\Fo$, generated by $s$ fermion  fields $\Psi_c(z)$, $c=1,...,s$ and define the phase space of the finite-dimensional spin CS system as the space of matrix coefficients 
	$$\pi_N\left(|v\rangle\right)=\langle 0|\UU(z_N)\UU(z_2)\cdots \UU(z_1)|v\rangle, \qquad |v\rangle\in\Fo$$
	where $\UU(z)=\sum_{c=1}^s \Psi_c(z)\otimes e_c$ and $e_c\in\C^s$ are basic vectors of $\C^s$. Then we systematically construct the pullback with respect to the maps $\pi_N$ of all operation required for the construction of the Yangian action on the finite-dimensional spin CS system. Finally this gives the Yangian action on the Fock space $\Fo$,
	which is a pullback of the Yangian action on finite-dimensional CS system. In particular, this  includes the construction of commuting Hamiltonians in $\Fo$.  Note the importance of the polynomial property of the total zero mode in the constructed Yangian action on the Fock space $\Fo$, which we do not prove. See Proposition \ref{proposition5}.
	
	The same language can be applied for the consideration of  bosonic limit of CS system constructed previously in \cite{KMS}. In this case the bosonic projection map 
	\beqq
\ppi_N\left(|v\rangle_{\!+}\right)={}_+\!\langle 0|\VV(z_N)\VV(z_2)\cdots \VV(z_1)|v\rangle_{\!+},\qquad |v\rangle_{\!+}\in\Ls
\eeqq
	actually coincides with that used in \cite{AMOS,AMY}. We include this case in Section 3, since it can clarify both the expositions of \cite{KMS} and of the main ideas of the present work given in Section 4.

	\section{Spin Calogero-Sutherland system}
The phase space of the quantum spin Calogero-Sutherland (CS) system consists of functions with values in vector space $\left(\C^{s}\right)^{\otimes N}$ while the dependence on spin in the Hamiltonian 
\begin{equation*}
H^{CS}= -\sum_{i=1}^{N}\left(\frac{\partial}{\partial q_i}\right)^2+\sum_{i,j=1}^N\frac{\beta(\beta-{K_{ij}})}{\sin^2(q_i-q_j)},
\end{equation*} is implicit \cite{KK}. 
Here $K_{ij}$ is the coordinate exchange operator of particles $i$ and $j$. After conjugating by the function $\prod_{i<j}|\sin(q_i-q_j)|^\beta$ which represents the degenerated vacuum state, and passing to the exponential variables $x_i=e^{2\pi i q_i}$ we arrive to the effective  Hamiltonian  
\begin{equation*}
H=\sum_{i=1}^{N}\left(x_i \frac{\partial}{\partial x_i}\right)^2+\beta\sum_{i<j}\frac{x_i+x_j}{x_i-x_j}\left(x_i\frac{\partial}{\partial x_i}-x_j\frac{\partial}{\partial x_j}\right)-2\beta\sum_{i<j}\frac{x_i x_j}{\left(x_i-x_j\right)^2}\left(1-K_{ij}\right).
\end{equation*}
which we restrict to the spaces $\L{N}_\pm$
of total invariants or respectively skewinvariants of the symmetric group $S_N$ in the space $V^{\otimes N}$,
\beqq
\L{N}_\pm=\left(V^{\otimes N}\right)^{(\pm)}.
\eeqq
Here\footnote{We also use the notation $V=V(z)$ when we need to specify the name of the variable} $V
=\C[z]\otimes \C^s$.
The (skew)invariants are taken with respect to the diagonal action of the symmetric groups, $\sigma_{ij}\mapsto K_{ij}P_{ij}$, where $K_{ij}$ is as above and $P_{ij}$ is the permutation of $i$-th and $j$-th tensor copy of the vector space $\C^s$. 

Further we use the Heckman--Dunkl operators $\D_i^{(N)}: V\otimes \L{N-1}_\pm\to  V\otimes \L{N-1}_\pm$ in the form suggested by Polychronakos \cite{Pol}:
\begin{equation} \label{Dunkl}
\D_{i}^{(N)}=x_i\frac{\partial}{\partial x_i}+\beta\sum_{j\neq i}\frac{x_i}{x_i-x_j}\left( 1 - K_{ij}\right).
\end{equation}
These operators  satisfy the relations
\begin{align}\nonumber
K_{ij}\D_i^{(N)}&=\D_j^{(N)} K_{ij}, \\ \nonumber
[\D_i^{(N)},\D_j^{(N)}]&=\beta (\D_j^{(N)}-\D_i^{(N)})K_{ij},
\end{align}
which coincide with the relations of the degenerate affine Hecke algebra after the renormalization  $\D_i^{(N)}\rightarrow \frac{1}{\beta}\D_i^{(N)}$.
By Drinfeld duality \cite{Dr}, this representation of degenerate affine Hecke algebra transforms to the 
representation of the Yangian  $\Y(\gl_s)$ in $\L{N}_\pm$, see \cite{Arakawa,KN} 
\begin{equation}\label{repYan}
t_{ab}(u)= \delta_{ab}+\beta\sum_i \frac{E_{ab,i}}{\beta u\pm\D_i^{(N)}} .
\end{equation}
Here $E_{ab,i}$ describes the action of $\gl_s$ on $i$-th tensor component,
$$
E_{ab,i}\Big(\dots\otimes\underbrace{(e^{c}\otimes x^{k})}_i\otimes\dots\Big)=\delta_{bc}\Big(\dots\otimes\underbrace{(e^{a}\otimes x^{k})}_i\otimes\dots\Big).
$$
and $t_{ab}(u)$, $a,b=1,\ldots s$, 
\beq\label{boson1} t_{ab}(u)=\delta^{ab}+\sum_{i=0}^{\infty}t_{ab,i}{u^{-i-1}}
\eeq
are  generating functions of the generators $t_{ab,i}$ of the Yangian $\Y(\gl_s)$. The defining relations of $\Y(\gl_s)$ are \cite{MNO}
\beq\label{???}
\left[t_{ab}(u),t_{cd}(v)\right]=\frac{t_{cb}(u)t_{ad}(v)-t_{cb}(v)t_{ad}(u)}{u-v}.
\eeq
Then the higher Hamiltonians of spin CS system can be chosen as coefficients 
of the quantum determinant
\begin{equation*} 
q\det t(u)= \sum_{\sigma\in S_m}(-1)^{sgn(\sigma)}
t_{\sigma(1),1}(u)t_{\sigma(2),2}(u-1) ... t_{\sigma(m),m}(u-m+1).
\end{equation*}
which generate the center of the $\Y(\gl_s)$ \cite{B,MNO}.

The main goal of the present paper is to construct the limit of the above Yangian action when $N$ tends to infinity. In particular, we get the limits of the above commuting family of Hamiltonians.

\setcounter{equation}{0}
\section{Bosonic limit}
In this section we observe the results of \cite{KMS} using slightly different language.

 Denote by $\Lambda^{(s)}$ the free unital associative commutative algebras generated by the elements
	$$\pk{a}, \qquad a=1,\ldots,s, \  k=1,2,\ldots $$
The ring $ \Lambda^{(s)}$ can be viewed as the ring of polysymmetric functions, that is the projective limit of polynomial functions on the variables 
$x_{11},\ldots x_{1n_1}, x_{21},\ldots x_{2n_2},\ldots x_{s1},\ldots x_{sn_s}$, symmetric on each group of variables $x_{a1},\ldots x_{an_a}$, $a=1,...,s$.
Here $\pk{a}$ corresponds to the Newton sums $x_{a1}^k+ x_{a2}^k+\cdots$.
 Denote by $\Ls$ the free unital associative commutative algebras generated by the elements
	$$\pk{a}, \qquad a=1,\ldots,s, \  k=0,1,2,\ldots $$ We have $\Ls\supset\Lambda^{(s)}$.
 Additional "zero modes" $p_{a,0}$ will further serve to count the numbers of variables in each group.

Let $\Heis{s}$ be the Heisenberg algebra with generators  $a_{c,k}, c=1,\ldots,s,\ k=0,1,...$ and
$\left(\q{c}\right)^{\pm 1}$, which satisfy the relations
\beq [a_{c,k},a_{d,l}]=k\delta_{cd}\delta_{k,-l},\qquad \q{c} a_{d,k}=(a_{d,k}+\delta_{cd}\delta_{k0})\q{c}.
\label{boson2}\eeq  
The space $\Ls$ is a representation of the Heisenberg algebra $\Heis{s}$, where
$$\begin{array}{cccc} 
a_{c,k}&\mapsto & p_{c,-k},& k\leq 0,\\
a_{c,k}&\mapsto &k\frac{\der}{\der p_{c,k}},& k>0,
\end{array},\qquad \q{c}\mapsto e^{\frac{\der}{\der p_{c,0}}}.$$
The unit of the ring $\Ls$ is then identified with the vacuum vector $|0\rangle_{\!+}$, so that 
\beq \label{boson3} a_{c,k}|0\rangle_{\!+}=0, \ c=1,...,s,\ \  k>0,\qquad \q{c}|0\rangle_{\!+}=|0\rangle_{\!+},\ c=1,...,s.\eeq 
Denote by ${}_+\!\langle 0|$ the vector of the dual space, which satisfies the relations
\beq \label{boson4}{}_+\!\langle 0|a_{c,k}=0,\qquad \ c=1,...,s,\ \ k\leq 0.
\eeq 
For $c=1,...,s$ denote by $\vf^-_c(z)$ the series
\beq\label{boson9} \vf^-_c(z)=\sum_{n\leq 0}a_{c,n}z^n\eeq
and by $e_c^\perp$ the linear operator $\C^s\to \C$ given by the relation 
$$e_c^\perp(e_b)=\delta_{bc}.$$
Define linear operators
\begin{align*}&\V{c}(z)=\exp\left(\sum_{n> 0}\frac{a_{c,n}}{n}z^n\right)\q{c}:\ &&\Ls\to \Ls\otimes\C[z], \qquad c=1,...,s, \\		
& \VV(z)=\sum_c \V{c}(z)\otimes e_c: && \Ls\to  \Ls\otimes V,\\
&\VVV(z)=\sum_c  \vf^-_c(z)\cdot\V{c}^{-1}(z)\otimes e_c^\perp:&& \Ls\otimes V\to \Ls\otimes  \C[z].
\end{align*}
For instance, for any $|v\rangle_{\!+}\in\Ls$
\beqq \VV^*(z) ( |v\rangle_{\!+}\otimes z^k \otimes e_c)=z^k\V{c}^{-1}(z)|v\rangle_{\!+},\eeqq
where
$$\V{c}^{-1}(z)= \vf^-_c(z)\cdot\q{c}^{-1}\exp\sum_{n> 0}-\frac{a_{c,n}}{n}z^n.$$

For any $|v\rangle_{\!+}\in\Ls$ consider the matrix element $\ppi_N(|v\rangle_{\!+})\in V^{\otimes N}$,
$$
\ppi_N(|v\rangle_{\!+})={}_+\!\langle 0|(\VV(z_N)\otimes 1^{\otimes(N-1)})\cdots(\VV(z_2)\otimes 1) \VV(z_1)|v\rangle_{\!+}
$$
which we shortly denote by
\beq\label{b1}
\ppi_N(|v\rangle_{\!+})={}_+\!\langle 0|\VV(z_N)\VV(z_2)\cdots \VV(z_1)|v\rangle_{\!+}
\eeq
 In components,
\beqq
\ppi_N(|v\rangle_{\!+})=\sum_{c_1,..,c_N=1}^s {}_+\!\langle 0|\V{c_N}(z_N)\cdots \V{c_1}(z_1)|v\rangle_{\!+}\cdot e_{c_N}\otimes \ldots \otimes e_{c_1}.
\eeqq
The commutativity 
\beq\label{b2} 
\V{b}(z_1)\V{c}(z_2)=\V{c}(z_2)\V{b}(z_1)
\eeq
implies that the matrix element \rf{b1} belongs to the space $\L{N}_+$.  Indeed,
	\begin{align*}
	\sigma_{ij}\left( \sum_{c_1,..,c_N=1}^s {}_+\!\langle 0|\cdots \V{c_j}(z_j)\cdots \V{c_i}(z_i)\cdots |v\rangle_{\!+}\cdot \ldots \otimes e_{c_j} \otimes\ldots\otimes e_{c_i}\otimes \ldots\right)=\\
	=\sum_{c_1,..,c_N=1}^s {}_+\!\langle 0|\cdots \V{c_j}(z_i)\cdots \V{c_i}(z_j)\cdots |v\rangle_{\!+}\cdot \ldots \otimes e_{c_i} \otimes\ldots\otimes e_{c_j}\otimes \ldots=\\
	=\sum_{c_1,..,c_N=1}^s {}_+\!\langle 0|\cdots \V{c_i}(z_i)\cdots \V{c_j}(z_j)\cdots |v\rangle_{\!+}\cdot \ldots \otimes e_{c_j} \otimes\ldots\otimes e_{c_i}\otimes \ldots
	\end{align*}
	In the last equality we change the indeces of summation $c_i$ by $c_j$.

Our goal is to  pull back the Yangian action \rf{repYan} in $\L{N}_+$ through the map $\ppi_N$.
The dissection of the relation \rf{repYan} shows that the application of each Yangian generator to a vector $|v\rangle_{\!+}\in\L{N}_+$ can be decomposed into several steps. 
First we present the symmetric tensor $|v\rangle_{\!+}\in\L{N}_+$ as an element of $\left(\C[x_i]\otimes \C^s\right)\otimes\L{N-1}_+$ for each tensor component, producing an equivariant family of vectors, which can be completely described by the element of
$V\otimes\L{N-1}_+\sim\left(\C[x_i]\otimes \C^s\right)\otimes\L{N-1}_+$ --- the decomposition of $|v\rangle_{\!+}$ over the first tensor component.
Then we apply the power of Heckman operator $\D_i^{(N)}$ to the $i$-th vector of this equivariant family and get another equivariant family. The last step is the symmetrization --- the sum of all members of the equivariant family.

Denote by $\iota_N:\L{N}_+\to V\otimes \L{N-1}_+$ the decomposition of the symmetric tensor $v$ over the first tensor component,
\beq\label{boson5}\iota_N\left(\sum_k f_{1k}(z)\otimes \cdots\otimes f_{Nk}(x_N)\right)=\sum_k f_{1k}(z)\otimes \left( f_{2k}(x_2)\otimes\cdots\otimes f_{Nk}(x_N)\right).\eeq
Here $f_{1k}(z)$ and $f_{jk}(x_k)$, $j>1$ are $\C^s$ valued polynomials.

\begin{lemma}\label{lemma1}. We have the following equality of linear maps $\Ls\to\L{N}_+$:
	\beq\label{boson7}
	\left( \ppi_{N-1}\otimes 1\right)\VV(z)=\iota_N\ppi_N.
	\eeq
\end{lemma}
{\bf Proof}. Applying both sides of \rf{boson7} to a vector $|v\rangle_{\!+}\in\Ls$ we get the tautology: both sides are equal to 
$${}_+\!\langle 0|\VV(x_N)\VV(x_{N-1})\cdots\VV(x_2) \VV(z)|v\rangle_{\!+}.$$
\hfill{$\square$}

For each tensor $u\in  V\otimes\L{N-1}_+$, symmetric with respect to diagonal permutations of all tensor factor except the first, denote by $E_N(u)$ its total symmetrization
\beq \label{boson13}E_N(u)=\sum_{j=1}^N \sigma_{1j}(u), \eeq
where $\sigma_{ij}=K_{ij}P_{ij}$ is the permutation of $i$-th and $j$-th tensor factors. On the other hand, for each $\F(z)\in  \Ls\otimes V$ define the element 
$\S ( \F(z))\in\Ls$  as the formal integral
\beq \label{boson11} \S ( \F(z))= \frac{1}{2\pi i}\oint \frac{dz}{z}\VVV(z)\F(z), \eeq
which counts zero term of the Laurent series. The following lemma establishe the map $\S$ as the pullback of the finite symmetrization. This is the crucial point of the construction.
\begin{lemma}\label{lemma2}
	For each $\F(z)\in\Ls\otimes  V$ and any natural $N$ we have the equality of elements of $\L{N}_+$:
	\beq\label{boson12}
	E_N( \ppi_{N-1}\otimes 1)(\F(z))=\ppi_N \S(\F(z)).
	\eeq
\end{lemma}
{\bf Proof}. Let $\F(z)$ has the form 
\beqq \F(z)=\sum_{c=1}^s F_c(z)\otimes e_c, \qquad F_c(z)\in \Ls\otimes \C[z].\eeqq
Consider first the LHS of \rf{boson12}.
This is the symmetrization \rf{boson13} of the tensor
\beqq
\sum_{c_1,..,c_N=1}^s {}_+\!\langle 0|\V{c_N}(x_N)\cdots \V{c_2}(x_2)F_{c_1}(x_1)\cdot e_{c_N}\otimes \ldots \otimes e_{c_1},
\eeqq
which can be written by means of proper changes of summation indices as the sum
\beqq
\sum_{k=1}^N\sum_{c_1,..,c_N=1}^s
{}_+\!\langle 0|\V{c_N}(x_N)\cdots \V{c_{k+1}}(x_{k+1})\V{c_{k-1}}(x_{k-1})\cdots\V{c_2}(x_2)\cdots F_{c_k}(x_k)\cdot e_{c_N}\otimes \ldots \otimes e_{c_1}.
\eeqq
Inserting in each summand the corresponding product
\beqq 1=\V{c_k}(x_k)\V{c_k}^{-1}(x_k)\eeqq
 and using the commutativity (\ref{b2}) we rewrite it as
\begin{align}\notag
&\sum_{k=1}^N\sum_{c_1,..,c_N=1}^s
{}_+\!\langle 0|\prod_{j=1}^N\V{c_j}(x_j)\cdot\V{c_k}^{-1}(x_k) F_{c_k}(x_k)\cdot e_{c_N}\otimes \ldots \otimes e_{c_1}=
\\
&\sum_{k=1}^N\sum_{c_1,..,c_N=1}^s
\frac{1}{2\pi i}\int_{z\around x_k}dz\ {}_+\!\langle 0|\prod_{j=1}^N\V{c_j}(x_j)\cdot\frac{\V{c_k}^{-1}(z)F_{c_k}(z)}{z-x_k}\cdot e_{c_N}\otimes \ldots \otimes e_{c_1}.
\label{boson14}\end{align}
The RHS of \rf{boson12} is 
\beqq \frac{1}{2\pi i}{}_+\!\langle 0|\prod_{j=1}^N\VV(x_j)\oint\frac{dz}{z}\VVV(z)\F(z). \eeqq
In components it looks as
\beq \label{boson14a}
\frac{1}{2\pi i}\sum_{a=1}^s\sum_{c_1,..,c_N=1}^s
{}_+\!\langle 0|\prod_{j=1}^N\V{c_j}(x_j)\oint\frac{dz}{z}{\vf_a^-(z)\V{a}^{-1}(z) F_{a}(z)}\cdot e_{c_N}\otimes \ldots \otimes e_{c_1}.
\eeq
The normal ordering of the above matrix elements assumes due to \rf{boson4} the move of
all $\vf_a^-(z)$ to the left vacuum using the relation
\beq
\V{c}(x)\vf_a^-(z)=\left(\vf_a^-(z)+
\frac{\delta_{ac}}{1-\frac{x}{z}}\right)\V{c}(x)\eeq
which follows from \rf{boson2}. In particular, the formal integral in \rf{boson14a} can be regarded as a contour integral, where the contour $C$ of integration encloses all points $x_j$. Since 
\beqq {}_+\!\langle 0|\vf_a^-(z) =0, \eeqq
we arrive to the expression
\beqq
\sum_{k=1}^N\sum_{c_1,..,c_N=1}^s
\frac{1}{2\pi i}\int_{z\around x_k}{}_+\!\langle 0|\prod_{j=1}^N\V{c_j}(x_j)\oint\frac{dz}{z}\frac{\V{c_k}^{-1}(z) F_{c_k}(z)}{1-\frac{x_k}{z}}\cdot e_{c_N}\otimes \ldots \otimes e_{c_1},
\eeqq
which is identical to \rf{boson14}. \hfill{$\square$}

We now apply statements of  Lemma \ref{lemma1} and \ref{lemma2}  for the construction of a  pullback of the Dunkl operator.   

Let $\D:\Ls\otimes V\to \Ls\otimes V$ be the linear map, such that for any $\F(z)\in\Ls\otimes V$
\beq \label{boson16}
\D \F^{(1)}(z)=z\frac{\dder}{\dder z}\F^{(1)}(z)+\frac{\beta z}{2\pi i}\oint \frac{d\xi}{\xi^2(1-\frac{z}{\xi})}\VVVV{2}(\xi)\VV^{(2)}(z)\F^{(1)}(\xi)
\eeq
Here the upper index $(i)$, $i=1,2$ indicates in which tensor copy of $\C^s$ the corresponding vector lives or an operator  acts. In components,
\beqq
\D(F_a(z)\otimes e_a)=\left( z\frac{\dder}{\dder z}F_a(z)+\frac{\beta z}{2\pi i}\oint \sum_{c=1}^s\frac{d\xi}{\xi^2(1-\frac{z}{\xi})}
\vf^-_c(\xi)\V{c}^{-1}(\xi)\V{c}(z)F_a(\xi)
\right)\otimes e_a 
\eeqq
We state that the operator $\D$ is the pullback of the equivariant family of Heckman operators $\D_i^{(N)}$.
\begin{proposition}\label{proposition1}
	For any $\F(z)\in\Ls\otimes V$ we have
	\beq\label{boson17}
	\left(\ppi_{N-1}\otimes 1 \right)\D(\F(x_1))=\D_1^{(N)}\left(\ppi_{N-1}\otimes 1 \right)\F(x_1)
	\eeq
\end{proposition}
{\bf Proof}. The only nontrivial part is the pullback of the difference part of the Heckman operator. 
The difference part $\tilde{\D}_{1}^{(N)}$ of Heckman operator $\D_1^{(N)}$ in the space $V(x_1)\otimes \L{N-1}_+$, where $V(x_1)=\C[x_1]\otimes \C^s$, can be described as the composition of three operations. First we include 
into $V(x_1)\otimes V(x_2)\otimes\L{N-2}_+$ by means of $1\otimes \iota_{N-1}$, then apply the operator $\dfrac{1-K_{12}}{x_1-x_2}$ and finally sum up over all the variables except $x_1$ by means of the summation $E_{N-1}$,
\beqq
\tilde{\D}_{1}^{(N)}=E_{N-1} \ \circ\ \dfrac{1-K_{12}}{x_1-x_2}\  \circ \
1\otimes \iota_{N-1}  
\eeqq 
The pullback of the inclusion $1\otimes \iota_{N-1}$ is $\,\VV^{(2)}(x_2)\,$ due to Lemma \ref{lemma1}, the pullback of  $E_{N-1}$ is $\ds\oint \VVVV{2}(x_2)\frac{dx_2}{x_2}$, the pullback of the operator $\dfrac{1-K_{12}}{x_1-x_2}$ is this very operator $\dfrac{1-K_{12}}{x_1-x_2}$. We see that the pullback of the difference operator 
$\tilde{\D}_{1}^{(N)}$ has the form
\beq\label{boson18}
\D \F^{(1)}(x_1)= \frac{x_1}{2\pi i}\oint \frac{dx_2}{x_2}\VVVV{2}(x_2)\frac{\VV^{(2)}(x_2)\F^{(1)}(x_1)-\VV^{(2)}(x_1)\F^{(1)}(x_2)}{x_1-x_2}
\eeq
Any matrix element of the ratio inside the integral is a polynomial on $x_1$ and $x_2$ and can be equally decomposed into a series either in the region $|x_1|<|x_2|$ or in the region $|x_1|>|x_2|$. In the region $|x_1|<|x_2|$ in the first integral
\beqq \frac{x_1}{2\pi i}\oint\frac{dx_2}{x_2}\VVVV{2}(x_2)\frac{\VV^{(2)}(x_2)\F^{(1)}(x_1)}{x_1-x_2}=\sum_{c=1}^s
\frac{x_1}{2\pi i}\oint\frac{dx_2}{x_2}\vf^-_c(x_2)\frac{\F^{(1)}(x_1)}{x_1-x_2}\eeqq
we have only negative powers of $x_2$ and this integral vanish. Thus we get \rf{boson16}. \hfill{$\square$}
\medskip

Let $E_{ab}\in \operatorname{End} \C^s$,  be the matrix unit, $E_{ab}(e_c)=\delta_{bc}e_a$. Denote by $\E_{ab}$, the operator $1\otimes 1\otimes E_{ab}:\Ls\otimes V\to \Ls\otimes V$: 
$$\E_{ab}\F(z)=F_b(z)\otimes e_a.$$ 
For $a,b=1,...,s$ and $n=1,...$ set
\beq \label{boson19}T_{ab,n}=\frac{(-1)^n}{2\pi i\beta^n}\oint \frac{dz}{z}\VVV(z)\E_{ab}\D^{n}\VV(z)
\eeq
Summarazing the statements above we get the following result \cite{KMS}
\begin{proposition}\label{proposition2} The operator $T_{ab,n}$,see \rf{boson19} is the pullback of the Yangian generator $t_{ab,n}$, see \rf{repYan},\rf{boson1}:
	\beqq \ppi_N T_{ab,n}= t_{ab,n} \ppi_N\qquad \text{for any}\qquad N\in\N.\eeqq 
\end{proposition}	
In particular, the operators \rf{boson19} form level zero representation of the Yangian $Y(\gl_s)$ in $\Ls$. Here we use the property 
\beq \label{boson20} \cap_{N\in\N} \operatorname{Ker}\,\ppi_N=0 \eeq
of the ring of symmetric functions which we assume to be known.

\setcounter{equation}{0}
\section{Fermionic limit}
Let $\HH$ be the algebra of $s$ free fermion fields. It is generated by the elements
$\psi_{nc}$ and $\psi^*_{nc}$, where $n\in\Z$ and $c=1,\ldots,s$, which subject the relations
\beq\label{ff1}\begin{split}
&\psi_{an}\psi_{bm}+\psi_{bm}\psi_{an}=0,\qquad 
\psi^*_{an}\psi^*_{bm}+\psi^*_{bm}\psi^*_{an}=0,\\
&\psi_{an}\psi^*_{bm}+\psi^*_{bm}\psi_{an}=\delta_{ab}\delta_{n,-m}.
\end{split}
\eeq
The algebra $\HH$ is graded with
\beq\label{ff7}
\deg \psi_{cn}=\deg\psi^*_{cn}=-n.
\eeq
 The algebra $\HH$ admits a family of commuting automorphisms
 $\Qoper_{c}$, $c=1,...,s$ given   by the relations
 \beq\label{ff5}\Qoper_{c}(\psi_{bn})=\psi_{b,n-\delta_{bc}},\qquad
 \Qoper_{c}(\psi^*_{bn})=\psi^*_{b,n+\delta_{bc}}.\eeq
Let $\Fo$ be the left representations of $\HH$, generated by the vacuum state $|0\rangle$,
and $\mathcal{F}^{s*}$ be the right  $\HH$ -module generated by the vacuume state 
$\langle 0|$, such that 
\beqq \langle 0|0\rangle=1 \eeqq
and
\beq\label{ff3}\begin{split} 
&\psi_{cn}|0\rangle= \psi_{cm}^*|0\rangle=0\qquad c=1,...,s,\qquad n\geq 0,\ m>0,\\
& \langle 0|\psi_{cn}=\langle 0|\psi_{cm}^*|=0, \qquad c=1,...,s,\qquad n< 0,\ m\leq 0.
\end{split}
\eeq 
 We use the following fermionic normal ordering rule:
\beq\label{normord}
 \vdots\psi_{cn}^*\psi_{dm}\vdots=\begin{cases}
\psi_{cn}^*\psi_{dm}, \ \ m\ge 0\\
-\psi_{dm}\psi_{cn}^*,\ \ m<0
\end{cases}.
\eeq
It is compatible with relations  (\ref{ff3}).

The automorphisms \rf{ff5} define  invertible  linear maps $\Q{c}$ and $\QQ{c}$ of the Fock space to itself which are compatible with these automorphisms  and anticommute for different indices $c_1$ and $c_2$:
\beq\label{ff6}\begin{split}
&\QQ{c}( x|0\rangle)=\Qoper_{c}^{-1}(x)\psi^*_{c,0}|0\rangle,\qquad \Q{c}( x|0\rangle)=\Qoper_{c}(x)\psi_{c,-1}|0\rangle,\\
&\langle 0|\QQ{c}= \langle 0|\psi_{c,1}^*,\qquad \qquad\qquad\qquad\langle 0|\Q{c}= \langle 0|\psi_{c,0},
\end{split}\eeq
so that for any $x\in\HH$ and $|v\rangle\in\Fo$ we have
\beq \label{fff6}\Qoper_c(x)|v\rangle=\Q{c} x \QQ{c}|v\rangle.
	\eeq
Indeed, for $|v\rangle=y|0\rangle$ the RHS of (\ref{fff6}) equals 
\beqq
\begin{split}
\Q{c} x \QQ{c}|v\rangle&=\Q{c} x \QQ{c}\left(y|0\rangle\right)=\Q{c}\left( x\Qoper_{c}^{-1}(y)\psi^*_{c,0}|0\rangle\right)=\Qoper_{c}(x)y\psi^*_{c,1}\psi_{c,-1}|0\rangle=\\
&=\Qoper_{c}(x)y\left(1-\psi_{c,-1}\psi^*_{c,1}\right)|0\rangle=\Qoper_{c}(x)y|0\rangle=\Qoper_c(x)|v\rangle.
\end{split}
\eeqq
In the following we use the distinguished product of $s$ such maps and automorphisms
\beq \label{ff6a}
\Qoper:=\Qoper_s\cdots\Qoper_1,\qquad \Q{}=\Q{s}\cdots \Q{1}.
\eeq
In particular,
\beq\label{ff6b}\begin{split} &\langle 0|\QQ{} = \langle 0|\psi_{s,1}^*\cdots\psi_{1,1}^*,\qquad\qquad
	\QQ{}|0\rangle = \psi_{s,0}^*\cdots \psi_{1,0}^*|0\rangle,\\
	& \langle 0|\Q{} = \langle 0|\psi_{s,0}\cdots\psi_{1,0},\qquad\qquad
	\Q{}|0\rangle = \psi_{s,-1}\cdots \psi_{1,-1}|0\rangle.
	\end{split}
\eeq
Denote by $\Hz$ the space of Laurent series 
\beq\label{H-z}
\sum_{n\in\Z}h_nz^n\in\Hz,
\eeq
where each coefficient $h_n$ is a
 series 
 \beqq h_n=\sum_ka^{11}_{k}+\sum_{kl}a^{21}_{k}a^{22}_{l}+\sum_{klm}a^{31}_{k}a^{32}_{l}a^{33}_{m}+\ldots\eeqq
 where $a^{ij}_k$ are either $\psi_{cn}$ or $\psi^*_{cn}$ for some $c$ and $n$, such that the matrix coefficient $\langle\xi|h_n|v\rangle$ is well  defined for any $\xi\in\Fstar$ and $v\in\Fo$. 
 \footnote{For instance the monomial $z^n\left(\sum_{k>0}\psi^*_{ck}\psi_{c,-k}\right)\notin \Hz$.} We can assume for instance all series $h_n$ to be fermionic normal ordered according to \rf{normord}, $h_n=\fnormal h_n\fnormal$.  We also use the notation $\Fo(z)$ for the space $\Fo\otimes\C[z,z^{-1}]]$.

Let $\U{c}(z)\ $ and $\U{c}^*(z)$ be the following elements of $\Hz$,
\beq\label{ff2}
\U{c}(z)=\sum_{n\in\Z}\psi_{cn}z^n,\qquad \U{c}^*(z)=\sum_{n\in\Z}\psi^*_{cn}z^{n-1}.
\eeq
The field $\U{c}(z)$ is of total degree zero, and the field $\U{c}^*(z)$ is of total degree $-1$, once we set $\deg z=1$.   
The relations \rf{ff3} imply the commutativity 
\beq \label{ff4a}\U{c}(x)\U{d}(y)+\U{d}(y)\U{c}(x)=\U{c}^*(x)\U{d}^*(y)+\U{d}^*(y)\U{c}^*(x)=0
\eeq
and normal ordering rules 
\beq\label{ff4}\begin{split}
&\U{c}(x)\U{d}(y)=\fnormal\U{c}(x)\U{d}(y)\fnormal, \qquad
\U{c}^*(x)\U{d}^*(y)=\fnormal\U{c}^*(x)\U{d}^*(y)\fnormal, \\
&\U{c}(x)\U{d}^*(y)=\fnormal\U{c}(x)\U{d}^*(y)\fnormal +\frac{\delta_{cd}}{y-x}, \qquad x<y,\\
&\U{c}^*(x)\U{d}(y)=\fnormal\U{c}^*(x)\U{d}(y)\fnormal +\frac{\delta_{cd}}{y-x} 
\qquad x<y.
\end{split}
\eeq
which imply the relation
 \begin{equation}\label{4}
\frac{1}{2\pi i}\int_{z\around w} \U{c}(w)\U{c}^*(z)dz =\frac{1}{2\pi i}\int_{z\around w} \U{c}^*(w)\U{c}(z)dz=1.
\end{equation} 
One can also see that
\beqq
\Qoper\left(\U{c}(z)\right)=z\U{c}(z),\qquad \Qoper\left(\U{c}^*(z)\right)=z^{-1}\U{c}^*(z),\qquad c=1,...,s.
\eeqq
Boson--fermion correspondence says that the space $\Fo$ is a representation of the affine Lie algebra $\widehat{\gl}_s$ of level one.
 The degree $-n$ generators $E_{ab,n}$ of $\widehat{\gl}_s$, where $a,b=1,...,s
 $ and $n\in\Z$ satisfy the relations
 \beqq 
[E_{ab,n},E_{cd,m}]=\delta_{bc}E_{ad,n+m}-\delta_{ad}E_{cb,n+m}+n\delta_{n,-m}\delta_{ad}\delta_{bc}\eeqq
 They are presented in $\End \Fo$ by operators
\beq \label{generaff} E_{ab,n}=\sum_{k+l=n}\fnormal \psi_{al}^*\psi_{bk}\fnormal, \eeq
where $\vdots \ \vdots $ means fermionic normal ordering (\ref{normord}).
 The generators 
 \beqq a_{b,n}:=E_{bb,n} \eeqq
 form the Heisenberg algebra $\H^s$,
 \beqq [a_{b,n},a_{c,m}]=n\delta_{b,c}\delta_{n,-m} \eeqq
 so that 
\beqq \label{ferm1} a_{c,k}|0\rangle=0,\qquad  
\langle 0|a_{c,l}=0, \qquad c=1,...,s,\ \ k\geq 0, l\leq 0.\eeqq
 On the other side of boson--fermion correspondence we have the relations:
 \begin{align*}&\U{c}(z)=z^{a_{c,0}}\exp\left(\sum_{n< 0}\frac{a_{c,n}}{n}z^n\right)\exp\left(\sum_{n> 0}\frac{a_{c,n}}{n}z^n\right)\Q{c}  \\
 &\U{c}^*(z)=z^{-a_{c,0}}\exp\left(-\sum_{n< 0}\frac{a_{c,n}}{n}z^n\right)\exp\left(-\sum_{n> 0}\frac{a_{c,n}}{n}z^n\right)\QQ{c}.			
 \end{align*}
The element
\beq \label{ferm3a}a_0=\sum_{c=1}^s a_{c,0}=\sum_{c=1}^s\sum_{k\in\Z} \vdots\psi_{ck}^*\psi_{c,-k}\vdots\eeq
is central in $\widehat{\gl}_s$ and satisfies the relation
\beqq \Q{}a_0\QQ{}=a_0+s.\eeqq
The Fock space $\Fo$  admits the orthogonal decomposition into direct sum of eigenspaces of operator $a_0$,
\beq \label{ferm3} \Fo=\oplus_{N\in\Z}\Fo_N,\qquad \text{where}\qquad \Fo_N=\{|v\rangle \in\Fo:\ a_0|v\rangle=N|v\rangle\}.
\eeq
The relation \rf{ferm3} implies that 
\beq\label{ferm4} \Q{}\Fo_N=\Fo_{N-s}.\eeq
In the following we use the notation $\tau_N$ for the projection of $\Fo$ to $\Fo_N$ parallel to other eigenspaces of $a_0$:
\beq\label{tau}
\tau_N|v\rangle)=\delta_{N,k}\cdot |v\rangle\ \  \text{     for   }\ \ |v\rangle\in\Fo_k.
\eeq

  Let $\UU(z)$ and $\UUU(z)$ be the following elements of $\Hz\otimes\Cs$ and $\Hz\otimes \Css$ correspondingly, 		
\begin{align} \label{ferm4a}	
	& \UU(z)=\sum_c \U{c}(z)\otimes e_c  		
 &&\UUU(z)=\sum_c \U{c}^{*}(z)\otimes e_c^\perp .
 \end{align}
 The field $\UU(z)$ defines the map from $\Fo$ to $\Fo(z)\otimes \Cs$,
 \beqq
 \UU(z)|v\rangle =\sum_c\U{c}(z)|v\rangle \otimes e_c,
 \eeqq
  which we denote by the same symbol $\UU(z)$. The field $\UUU(w)$ defines a map from $\Hz\otimes \Cs$ to $\H_-^s(z,w)$,
  \beqq 
  \UUU(w) \left(\sum _c F_c(z)\otimes e_c\right) =\sum_c\U{c}^*(w)F_c(z), \eeqq 
  where  $\H_-^s(z,w)$ is defined in the same way as $\Hz$ (\ref{H-z}).
  Here we regard $e_c^\perp$ as the linear map $e_c^\perp:\C^s\to\C$, such that $e_c^\perp(e_d)=\delta_{cd}$.
  	
 For any $|v\rangle\in\Fo$ consider the matrix element
 $$
 \pi_N(|v\rangle)=\langle 0|(\UU(z_N)\otimes 1^{\otimes(N-1)})\cdots(\UU(z_2)\otimes 1) \UU(z_1)|v\rangle,
 $$
 which we shortly denote by
 \beq\label{ferm6}
 \pi_N(|v\rangle)=\langle 0|\UU(z_N)\UU(z_2)\cdots \UU(z_1)|v\rangle.
 \eeq
 In components,
 \beqq
 \pi_N(|v\rangle)=\sum_{c_1,..,c_N=1}^s \langle 0|\U{c_N}(z_N)\cdots \U{c_1}(z_1)|v\rangle\cdot e_{c_1}\otimes \ldots \otimes e_{c_N}.
 \eeqq
  The commutativity \rf{ff4a} and the properties of the left vacuum \rf{ff3}
 imply that the matrix element \rf{ferm6} belongs to the space $\L{N}_-$.
  Note that the map $\pi_N$ factors through the projection $\tau_N$ (\ref{tau}),
  $$\pi_N=\pi_N\tau_N$$
  and equals zero for any $\Fo_\lambda$ with $|\lambda|\not=N$.
  
   Analogously to the rings of symmetric functions in $N$ variables and to the spaces of antisymmetric functions in $N$ variables, the spaces $\L{N}_-$ form a projective system.  
  Regard an element $f$ of $\L{N+s}$ as $\left(\Cs\right)^{\otimes(N+s)}$ valued function $f=f(x_1,x_2,\ldots, x_{N+s})$. Set
  \beq\label{ferm6a} \omega_N(f)=\frac{\left(1^{\otimes N}\otimes e_1^{\perp}\otimes e_2^{\perp}\cdots\otimes  e_s^{\perp}\right)
  f(x_1,\ldots, x_N,0,\ldots 0)}{x_1\cdots x_N}.\eeq
  In components,
  \beqq\begin{split} \omega_N\!\left(x_1^{a_1}e_{c_1}\!\otimes\ldots\otimes x_{N+s}^{a_{N+s}}e_{c_{N+s}}\right)=
  	\delta_{a_{N+1,0}}\cdots \delta_{a_{N+s,0}}\!\delta_{c_{N+1},1}\!\cdots \delta_{c_{N+s},s}
  & x_1^{a_1-1}\!e_{c_1}\otimes\ldots\otimes x_{N}^{a_{N}-1}\!e_{c_{N}}.
\end{split}\eeqq
 One can see that $\omega_N$ is a linear map from $\L{N+s}$ to $\L{N}$. 
 \begin{lemma} \label{lemma0} For any $|v\rangle\in\Fo$ we have the equality
 	\beq\label{ferm7}
 	\pi_N(\Q{}|v\rangle)=\omega_N\cdot \pi_{N+s}(|v\rangle)	.
 	\eeq 	
 	\end{lemma}
 {\bf Proof}. The LHS of \rf{ferm7} reads
 \beqq \begin{split} \pi_N(\Q{}|v\rangle)=&\langle 0|\UU(x_N)\cdots \UU(x_2)\UU(x_1)\Q{}|v\rangle=\\
 	&(x_1\cdots x_N)^{-1}\cdot
 \langle 0|\Q{}\UU(x_N)\cdots\UU(x_2)\UU(x_1)|v\rangle=
 \\
 &(x_1\cdots x_N)^{-1}\cdot
 \langle 0|\psi_{s,0}\cdots\psi_{1,0} \UU(x_N)\cdots\UU(x_2)\UU(x_1)|v\rangle.
\end{split} \eeqq
The last line is precisely the RHS of \rf{ferm7}.
Indeed,
\beqq
\begin{split}
&\omega_N\cdot \pi_{N+s}(|v\rangle)=\omega_N\cdot\sum_{c_1,..,c_{N+s}=1}^s \langle 0|\U{c_{N+s}}(x_{N+s})\cdots \U{c_2}(x_2)\U{c_1}(x_1)|v\rangle\cdot e_{c_1}\otimes \ldots \otimes e_{c_{N+s}}\\
&=(x_1\cdots x_N)^{-1}\cdot\sum_{c_1,..,c_{N}=1}^s \langle 0|\psi_{s,0}\cdots\psi_{1,0}\U{c_{N}}(x_{N})\cdots \U{c_2}(x_2)\U{c_1}(x_1)|v\rangle\cdot e_{c_1}\otimes \ldots \otimes e_{c_{N}}.
\end{split}
\eeqq
\hfill{$\square$}

  We are going now to construct the pullback through the maps $\pi_N$ of the components of the Yangian generators.
  
  Denote by $\iota_N:\L{N}_-\to \Cs[z]\otimes \L{N-1}_-$ the decomposition of the antisymmetric tensor $v$ over the first tensor component, given by the relation \rf{boson5}.
  Denote by $\pi_{N-1,1}: \left(\Hz\otimes \C^s\right)\otimes\Fo\to \C[x_2,...,x_N,z,z^{-1}]\otimes{\C^s}^{\otimes N}$ the map defined as
  \beqq \pi_{N-1,1}(\F(z)\otimes |v\rangle)=
  \langle 0|\UU(x_N)\cdots\UU(x_2) \F(z)|v\rangle. \eeqq
  \begin{lemma}\label{lemma3}. We have the following equality of linear maps $\Fo\to\L{N}_-$:
  	\beq\label{ferm9}
  	 \pi_{N-1,1}\UU(z)=\iota_N\pi_N.
  	\eeq
  \end{lemma}
{\bf Proof}. This is again a tautology like in the proof of Lemma \ref{lemma1}.\hfill{$\square$}

For each polynomial tensor $\Cs[z]\in  V\otimes\L{N-1}_-$, antisymmetric with respect to diagonal permutations of all tensor factor except the first, denote by $\Alt_N(u)$ its total (nonnormalized) antisymmetrization
\beq \label{ferm10}\Alt_N(u)=u-\sum_{j=2}^N \sigma_{1j}(u). \eeq
 On the other hand, for each $\F(z)\in  \Hz\otimes \C^s$ define the element 
$\A ( \F(z))\in\H$   as the integral
\beq \label{ferm11} \A ( \F(z))= 
 \frac{1}{(2\pi i)^2}\int_{z\around 0}dz\int_{u\around z}du\frac{\UUU(u)\F(z)}{u-z} .\eeq
{\bf Remark}. The integral over $z$ is actually formal. The form \rf{ferm11} indicates the following. Assume that $\F(z)$ depends on a parameter $w$. Then  the contour $C$ of integration over $z$ should not enclose the point $z=w$. One can always assume the condition $|w|>|z|$.
\medskip

Let an element $\F(z)\in \Hz\otimes \C^s$ satisfies the following conditions:
\begin{align} \label{ferm12}
&(i)&& \pi_{N,1}(\F(z)\otimes |v\rangle) &&\text{is a polynomial on} \ z \ \text{for any} \ N\in\N,\ v\in\Fo\\
&(ii)&&\deg\F(z)=0 \label{ferm13}
\end{align}
Here we assume that $\deg e_c =0$  for any $e_c\in\C^s$. 
 
The following lemma establishes the map $\A$ as the pullback of the finite antisymmetrization. This is the crucial point of the construction.
\begin{lemma}\label{lemma4}
	For each $\F(z)\in \Hz\otimes \C^s$ satisfying the conditions \rf{ferm12} and \rf{ferm13}, any $|v\rangle\in\Fo$ and any natural $N$ we have the equality of elements of $\L{N}_-$:
	\beq\label{ferm14}
	\Alt_N \pi_{N-1,1}(\F(z)\otimes|v\rangle)=\pi_N \A(\F(z))|v\rangle.
	\eeq
\end{lemma}
{\bf Proof}.  Let $\F(z)$ has the form 
\beqq \F(z)=\sum_{c=1}^s F_c(z)\otimes e_c, \qquad F_c(z)\in \Hz\eeqq
Consider first the LHS of \rf{ferm14}.
This is the antisymetrization \rf{ferm10} of the tensor
\beqq
\sum_{c_1,..,c_N=1}^s \langle 0|\U{c_N}(x_N)\cdots \U{c_2}(x_2)F_{c_1}(x_1)|v\rangle\cdot e_{c_1}\otimes \ldots \otimes e_{c_N}.
\eeqq
which can be written by means of proper changes of summation indices as the sum
\beqq\begin{split}
\sum_{k=1}^N(-1)^{k+1}\!\!\!\! \sum_{c_1,..,c_N=1}^s
\langle 0|\U{c_N}(x_N)\cdots \U{c_{k+1}}(x_{k+1})\U{c_{k-1}}(x_{k-1})& \cdots\U{c_1}(x_1) F_{c_k}(x_k)|v\rangle\\
&\cdot e_{c_1}\otimes \ldots \otimes e_{c_N}.
\end{split}\eeqq
Using the relation  $$\displaystyle \int_{z\around x_k}dz\ \frac{ F_{c_k}(z)}{x_k-z}=-F_{c_k}(x_k),$$ we rewrite the LHS of \rf{ferm14} as
\beqq\begin{split}
\sum_{k=1}^N(-1)^{k}\!\!\!\! \sum_{c_1,..,c_N=1}^s
\langle 0|\U{c_N}(x_N)\cdots \U{c_{k+1}}(x_{k+1})\U{c_{k-1}}(x_{k-1})\cdots\U{c_1}(x_1) \cdot\\
\frac{1}{2\pi i}\int_{z\around x_k}dz\frac{ F_{c_k}(z)}{x_k-z}|v\rangle\cdot e_{c_1}\otimes \ldots \otimes e_{c_N}.
\end{split}\eeqq 
 Using \rf{4}, we insert the integral:
 $$\frac{-1}{2\pi i}\int_{u\around x_k} \U{c_k}(x_k)\U{u}^*(u)du=1$$
into each summand of the $k$-th group. Then the LHS of \rf{ferm14} takes the form 
 \begin{align}\notag
 &-\frac{1}{(2\pi i)^2}\sum_{k=1}^N\! \sum_{c_1,..,c_N=1}^s\int\limits_{z\around x_k}dz\!\!\!\!\!\!\!\int\limits_{\substack{u\around x_k\\|z-x_k|\gg|u-x_k|}}\!\!\!\!\!\!\!\!\!\!du\ \langle 0|\prod_{N\geq i\geq 1}\!\U{c_i}(x_i) \frac{\U{c_k}^*(u)F_{c_k}(z)}{x_k-z}|v\rangle 
 \cdot  e_{c_1}\!\otimes\! \ldots\! \otimes e_{c_N}=	\\ \label{ferm15}
 &-\frac{1}{(2\pi i)^2}\sum_{k=1}^N\! \sum_{c_1,..,c_N=1}^s\int\limits_{z\around x_k}dz\!\!\!\!\!\!\!\int\limits_{\substack{u\around x_k\\|z-x_k|\gg|u-x_k|}}\!\!\!\!\!\!\!\!\!du\ \langle 0|\prod_{N\geq i\geq 1}\!\U{c_i}(x_i) \frac{\U{c_k}^*(u)F_{c_k}(z)}{u-z} |v\rangle
 \cdot  e_{c_1}\!\otimes \!\ldots \!\otimes e_{c_N}.
 \end{align}
  Now in each summand we move the contour of integration for $z$ close to the point $x_k$, crossing the singularity at $z=u$. Then the integral in every such summand transforms into the sum of two integrals, 
  \begin{align}\notag &-\int\limits_{z\around x_k}dz\!\!\!\!\!\!\!\int\limits_{\substack{u\around x_k\\|z-x_k|\gg|u-x_k|}}\!\!\!\!\!\!\!\!\!du\ \langle 0|\prod_{N\geq i\geq 1}\!\U{c_i}(x_i) \frac{\U{c_k}^*(u)F_{c_k}(z)}{u-z}|v\rangle= \\
  \label{ferm16} &-\int\limits_{z\around x_k}dz\!\!\!\!\!\!\int\limits_{\substack{u\around x_k\\|u-x_k|\gg|z-x_k|}}\!\!\!\!\!\!\!\!\!du\ \langle 0|\prod_{N\geq i\geq 1}\!\U{c_i}(x_i) \frac{\U{c_k}^*(u)F_{c_k}(z)}{u-z}|v\rangle+\\
   \notag &\int\limits_{z\around x_k}dz\int\limits_{{u\around z}}du\ \langle 0|\prod_{N\geq i\geq 1}\!\U{c_i}(x_i) \frac{\U{c_k}^*(u)F_{c_k}(z)}{u-z}|v\rangle
  \end{align}
  In the first integral, see the middle line of \rf{ferm16}, after the change of the order of integration we observe its vanishing due to condition $(i)$ of \rf{ferm12}: there is no singularity of the integral at 
  any point $z=x_j$. We now conclude that the LHS of \rf{ferm14} equals to the double integral
  \beqq
  \frac{1}{(2\pi i)^2}\sum_{k=1}^N\! \sum_{c_1,..,c_N=1}^s\int\limits_{z\around x_k}dz\!\int\limits_{{u\around z}}\!du \langle 0|\prod_{N\geq i\geq 1}\!\U{c_i}(x_i) \frac{\U{c_k}^*(u)F_{c_k}(z)}{u-z} |v\rangle
  \cdot  e_{c_1}\!\otimes \ldots \otimes e_{c_N}
  \eeqq
  or 
   \beq\label{ferm17}
  \frac{1}{(2\pi i)^2}\sum_{k=1}^N\! \sum_{c_1,..,c_N=1}^s\int\limits_{ C}dz\!\int\limits_{{u\around z}}\!du \langle 0|\prod_{N\geq i\geq 1}\!\U{c_i}(x_i) \frac{\U{c_k}^*(u)|F_{c_k}(z)}{u-z} |v\rangle
  \cdot  e_{c_1}\!\otimes \ldots \otimes e_{c_N},
  \eeq
   where the contour $C$ encloses all the points $x_k$ but does not enclose zero.
   \medskip
   
 On the other hand,  
 the RHS of \rf{ferm14},
 \beqq
 \langle 0|\frac{1}{(2\pi i)^2}\prod_{N\geq i\geq 1}\UU(x_i)\oint dz\int_{u\around z}du\frac{\UUU(u)\F(z)}{u-z}|v\rangle
 \eeqq
 in components looks like
 \beqq
 \frac{1}{(2\pi i)^2}\sum_{k=1}^N\! \sum_{c_1,..,c_N=1}^s\langle 0|
 \prod_{N\geq i\geq 1}\U{c_i}(x_i)\oint dz\int_{u\around z}du\ \frac{\U{c_k}^*(u)F_{c_k}(z)}{u-z} |v\rangle
 \cdot  e_{c_1}\otimes \ldots \otimes e_{c_N}.
 \eeqq
 The region of analyticity of any matrix coefficient $\langle \xi|\prod_{N\geq i\geq 1}\U{c_i}(x_i)\U{c_k}^*(u)F(z)|v\rangle\ $ is \ $0<|x_N|<|x_{N-1}|<\ldots<|x_1|<|u|<|z|$ so the  integral over $z$ can be replaced by the contour integral over the contour enclosing
 all $x_k$ and zero. Deforming this contour we see that the RHS of \rf{ferm14} equals to the sum \rf{ferm15} plus the integral which enclose zero and not the points $x_k$.
  To prove the equality \rf{ferm14} it is sufficient to verify that each
   integral
   \beq \label{ferm18}
  \int\limits_{z\around 0}dz\int\limits_{u\around z}du\ \langle 0|\prod_{N\geq i\geq 1}\!\U{c_i}(x_i) \frac{\U{c_k}^*(u)F_{c_k}(z)}{u-z} |v\rangle
   \eeq
    vanishes. 
  In the latter integral all singularities  at the diagonals $z=x_i$ and $u=x_i$ are out of the domain of integration.  
  Thus the vanishing of these integrals is equivalent to vanishing of the vector valued integral
  \beq\label{ferm20}
  \int\limits_{z\around 0}dz\int\limits_{u\around z}du\ \langle 0|\frac{\U{c_k}^*(u)F_{c_k}(z)}{u-z}.
  \eeq
  The domain of analyticity of the expression $\U{c_k}^*(u)F_{c_k}(z)$ is $|u|<|z|$ and all the singularities are poles of finite order on the diagonal $z=u$, so that the relation
  \begin{equation}\label{c1}
  (z-u)^N \U{c_k}^*(u)F_{c_k}(z)=(z-u)^N F_{c_k}(z)\U{c_k}^*(u)
  \end{equation}
  holds for sufficiently big $N$, where the sign is chosen according to parity of the field  $F_{c_k}(z)$.
  The relation (\ref{c1}) implies that the analytic continuation of
  $\U{c_k}^*(u)F_{c_k}(z)$ to the region $|u|>|z|$ is $\pm F_{c_k}(z)\U{c_k}^*(u)$.
   By definition \rf{ff3} of the vacuum state and the related rules of the normal ordering  (\ref{normord}) the integral \rf{ferm20} can be
    formally rewritten as
    \beq \label{ferm20a}\int_{z\around 0}dz\langle 0|\left(\Psi_{c_k,-}^*(z) F_{c_k}(z)\pm F_{c_k}(z)\Psi_{c_k,+}^*(z)\right),
    \eeq
    where the sign depends on the parity of $F_{c_k}(z)=\sum_{n\in\Z}f_nz^n$ and
    \beqq \Psi_{c_k,-}^*(z)=\sum_{n\leq 0}\psi_{c_k,n}^*z^{n-1},\qquad
    \Psi_{c_k,+}^*(z)=\sum_{n> 0}\psi_{c_k,n}^*z^{n-1}. \eeqq
    In Fourier modes \rf{ferm20} looks as
    \beqq \sum_{n\leq 0} \psi_{c_k,n}^* f_{-n}\pm \sum_{n>0} f_{-n} \psi_{c_k,n}^*.
    \eeqq
    The first sum vanishes due to \rf{ff3}. By assumption, $\deg F_{c_k}(z)=0$ thus $\deg f_n=-n$. We then see that
     in the second term all $f_{n}$ have positive degree and  if we assume them to be normal ordered they 
     contain in each summand either $\psi_{a n}$ or $\psi^*_{bn}$ with $n<0$ at their left end. Thus $\langle 0|f_{-n}=0$
      for $n>0$ and the integral \rf{ferm20} vanishes. \hfill{$\square$}

  Define an operator $\D:\Hz\otimes\Cs\to \Hz\otimes\Cs$ by the relation    
   \beq \label{ferm19}\begin{split}
  &\D \F^{}(z)=z\frac{\dder}{\dder z}\F^{}(z)+\\&\frac{\beta z}{(2\pi i)^2}\int_{\substack{w\around 0\\ |w|<|z|}}dw\int_{u\around w}du\ \UUUU{2}(u) \frac{\UU^{(2)}(w)\F^{(1)}(z)-\UU^{(2)}(z)\F^{(1)}(w)}{(u-w)(z-w)}.
  \end{split}\eeq 
Here upper indices $(1)$ and $(2)$ indicate tensor components where corresponding operators act. In components,
 \beqq \begin{split} &\D F_c(z)\otimes e_c=  z\frac{\dder}{\dder z}F_c^{}(z)\otimes e_c+\\ &\frac{\beta z}{(2\pi i)^2}
 	\sum_{b=1}^s \int_{\substack{w\around 0\\ |w|<|z|}}dw\int_{u\around w}du\ \U{b}^*(u)
 	\frac{\U{b}(w)F_c(z)-\U{b}(z)F_c{}(w)}{(u-w)(z-w)}\otimes e_c.
\end{split}\eeqq

By means of of Lemma \ref{lemma4} we now can identify the operator $\D$ as a pullback of the equivariant family of Heckman operators $\D_i^{(N)}$ acting in the space of partially antisymmetric tensors
\begin{proposition}\label{proposition3}
	For any $\F(z)\in \Hz\otimes\Cs$ satisfying the condition \rf{ferm12} and \rf{ferm13}, $|v\rangle\in\Fo$ and $N\in\N$
	 we have the equality
	\beq\label{ferm21}
	\pi_{N-1,1}(\D\F(x_1)\otimes|v\rangle)=\D_1^{(N)}\pi_{N-1,1}(F(x_1)\otimes|v\rangle).
	\eeq
\end{proposition}
{\bf Proof}. First we note that once the element $\F(z)\in \Hz\otimes\Cs$ satisfies the conditions \rf{ferm12} and \rf{ferm13}, the same is true for the divided difference
\beqq \frac{\UU^{(2)}(w)\F^{(1)}(z)-\UU^{(2)}(z)\F^{(1)}(w)}{z-w}.\eeqq
 The property \rf{ferm12} is valid because both the differential and difference derivatives preserve the polynomial property. The property \rf{ferm13} is evident: the difference derivatives are homogeneous of degree zero.
  We thus can use Lemma \ref{lemma4}. Now the rest of the proof is identical to  the  
  proof of Proposition \ref{proposition1}. \hfill{$\square$}
\medskip

Note that the application of the operator $\D$ to some $\F(z)\in \Hz\otimes\Cs$, which satisfies the conditions \rf{ferm12} and \rf{ferm13},
preserves these conditions by the same reasons of homogeneity and preservation of polynomial spaces by both difference and differential derivatives. 
This gives rise to the formulas for pullback of sum of powers of Dunkl operators.

 Let $E_{ab}\in \operatorname{End} \C^s$,  be the matrix unit, $E_{ab}(e_c)=\delta_{bc}e_a$. Denote by $\E_{ab}$, the operator $ 1\otimes E_{ab}:\Hz\otimes \Cs\to \Hz\otimes \Cs$: 
$$\E_{ab}\F(z)=F_b(z)\otimes e_a.$$ 
For $a,b=1,...,s$ and $n=1,...$ define the element $T_{ab,n}\in\HH$ by the relation
\beq \label{ferm22}T_{ab,n}=\beta^{-n}\A \E_{ab}\D^{n}\UU(z)=
\frac{1}{2\pi i}\int_{z\around 0}dz\ \T_{ab,n}.
\eeq
Here $\T_{ab,n}$ is the $n$-th order density defined by the formula:
\beq\label{dens}
\T_{ab,n}=\frac{1}{2\pi i \beta^n}\int_{u\around z}du\frac{\UUU(u)\E_{ab}\D^{n}\UU(z)}{u-z}.
\eeq
In Appendix we give expressions for the first densities for $n=0, 1, 2$ in terms of normal ordering fermionic fields and in terms of generators of the affine Lie algebra $\widehat{\gl}_s$ , see (\ref{generaff}).

Summarizing the statements above we establish the operator $T_{ab,n}$ as the pullback of the Yangian generator $t_{ab,n}$ in $\Lambda^{s,N}_-$. 
\begin{proposition}\label{proposition4} For any $|v\rangle\in\Fo$ and $N\in\N$ we have the equality 
	\beqq \pi_N (T_{ab,n}|v\rangle)= t_{ab,n} \pi_N|v\rangle.\eeqq
\end{proposition}
 Moreover these operators themselves form a representation of the Yangian $Y(\gl_s)$:
 \begin{proposition}\label{proposition5} The operators $T_{ab,n}$ satisfy Yangian relations \rf{???}
 	\end{proposition}
In particular, the coefficients of the quantum determinant $q \det T(u)$ form a commutative family which can be regarded as the limits of the higher Hamiltonians of CS system.
 	
  Contrary to the bosonic case, this statement is not straightforward since the intersections of the kernels of maps
 $\pi_N$ is not zero. For instance all subspaces $\Fo_N$ with $N<0$ belong to this kernel.  Below we present  arguments in favour
  of the statement which can be developed up to a rigorous proof.
 
 Recall  the characterization of
 the subspace $\Fo_N$
  as the eigenspace of the operator $a_{0}$, see \rf{ferm3}:
  \beqq \Fo_N\ =\ \{|v\rangle:\ a_0 |v\rangle\,= N |v\rangle.\} \eeqq
 The space $\Fo_N$ contains the vector $|N\rangle= \Q{}^{-N}|0\rangle$ and any vector $|v\rangle\in\Fo_N$ can be obtained from
   $|N\rangle$ by applying some element of zero charge,
  \beqq| v\rangle= x|N\rangle,\qquad\text{where}\qquad \Q{}x\QQ{}=x,\eeqq
 for instance $x$ is a sum of products of fermions where in each monomial the  total number of $\psi_{cn}$ equals to the total number
  of $\psi_{cn}^*$. Alternatively, any vector $|v\rangle\in\Fo_N$ can be obtained from
  $|N\rangle$ by applying elements of affine Lie algebra $\widehat{\gl}_s$. In this realization powers of the operator $\Q{}$ identify different spaces $\Fo_N$,
  \beqq |v\rangle= x|N\rangle \Longrightarrow \Q{}^m|v\rangle=x|N\rangle\qquad \text{if}\qquad \Q{}x\QQ{}=x. 
  \eeqq
 The crucial property of the constructed operators $T_{ab,n}$, which force them to preserve the Yangian relations, is their {\it polynomial} dependence on $a_0$. Namely, each operator $A=T_{ab,0}$ can be presented as a polynomial
 \beq\label{ferm23} A=A_0+a_0A_1+\ldots+a_0^kA_k,\qquad\Q{}A_j\QQ{}=A_j \eeq
where each $A_j$ is an element of $\H$ of zero charge presenting an operator in $\Fo$. Indeed, each Yangian relation became nontrivial in the space $\L{N}$ for big enough $N$. This means that for any vector
$|v\rangle\in\Fo$ this relation is nontrivial on $\pi_{N}(\Q{}^{k+N}|v\rangle)$ for some $k$ and big enough $N$ and thus any matrix coefficient of the corresponding relation in the Heisenberg algebra is a polynomial on $N$ equal zero for $N$ big enough and thus is identically zero.

The element $T_{ab,n}$ is by construction is an ordered  series in elements of $\HH$ of total degree zero in a sense of \rf{ff5}, with each summand being a product of at most $2n$ elements $\psi_{ck}$ and $n$ elements $\psi_{bm}^*$. Calculation for small $n$ show that it can be rewritten as an ordered series of elements of $\widehat{\gl}_s$ with each summand be a product of at most $n+1$ generators $E_{ab,m}$ of $\widehat{\gl}_s$. Thus $T_{ab,n}$ should be a polynomial on $a_0$ of order $n+1$. 
 
  We have no rigorous proof of the latter statement. Instead  one can equivalently establish the relation
  \beq\label{ferm24} \operatorname{ad}_Q^{n+1}( T_{ab,n})=0\eeq
  where $\operatorname{ad}_Q(x)=Qx-xQ$, and check this equality on $\Fo_N$ for big $N$. The equality \rf{ferm24} is evidently clear for the differential part of sums of power of Dunkl operators,
  $$A=\fnormal \sum_{k,c}k^n\psi_{c,-k}^*\psi_{ck}\fnormal$$ 
  Indeed, 
  \beq\label{ferm25}\Q{}A\QQ{}-\D_n=\sum_{k,c}\left((k+1)^n-k^n \right)\fnormal \psi_{c,-k}^*\psi_{ck}\fnormal.\eeq
  The coefficients at the RHS are now polynomials on $k$ of order $n-1$. Repeating \rf{ferm25} we get the equality $\operatorname{ad}_Q^{n+1}(A)=0$.
  
  Another way to check relations \rf{ferm24} is to use the realization of the operator $\Q{}$ in antisymmetric functions of finite variables, see Lemma \ref{lemma0}. Namely, denote by $\omega_{N+1,N+s}^\perp$ the operator $e_1^\perp\otimes\ldots \otimes e_s^\perp$ applied to the last tensor components of the space $\L{N+s}$, see \rf{ferm6a}. 
  
  Consider an operator  $\AA=\sum\AA_i$ in $\L{N}$  being the sum of powers of Dunkl operators multiplied possibly by $E_{ab}$. Let  $A$ be its 
  pullback to the space $\Fo$ given by the relation \rf{ferm22}. Choose a vector $|v\rangle\in\Fo_{N+s}$. Let $f(x_1,\ldots x_{N+s})=\pi_{N+s}(|v\rangle)$ be the corresponding vector valued function.     Then by Lemma \ref{lemma0} the difference 
  $ \pi_N(\Q{} A -A \Q{} )|v\rangle$ can be presented as a sum
  \beq \label{ferm26}\begin{split}\pi_N(\Q{} A -A \Q{})|v\rangle= \omega_{N+1,N+s}^\perp\prod_{k=1}^N x_k^{-1}\sum_{i=1}^{N+s}\left(\AA_if(x_1,\ldots,x_{N+s})\right)|_{x_{N+1}=\ldots= x_{N+s}=0}-\\
  \sum_{i=1}^{N}\AA_i\left(\omega_{N+1,N+s}^\perp\prod_{k=1}^N x_k^{-1}f(x_1,\ldots,x_{N+s})|_{x_{N+1}=\ldots= x_{N+s}=0}    \right)	\end{split}
  \eeq
  We are to show that finite iteration of the RHS of \rf{ferm26} vanishes. The nontrivial part is to verify this for sums of powers of Dunkl operators. Note that due to the structure of Dunkl operators both sums in RHS of \rf{ferm26} can be taken from $1$ to $N$.  For instance, if $\AA_i$ is a differentiation, $\AA_i=x_i\frac{\partial}{\partial x_i}$, their sum measure degree of homogeneity, which differs by $N$ for two summands of \rf{ferm26}.
  Hence the Euler operator has order two in accordance to what we showed before. If $\AA$ is a sum of difference parts of Dunkl operators, then  the RHS can be reduced to the form   
  \begin{align*} & \omega_{N+1,N+s}^\perp\prod_{k=1}^N x_k^{-1}\sum_{i=1}^{N}\sum_{j=N+1}^{N+s}\left(\sum_{i=1}^Nx_i\frac{(1-K_{ij})f}{x_i-x_j}
  \right)|_{x_{N+1}=\ldots= x_{N+s}=0}=\\
   & \omega_{N+1,N+s}^\perp\prod_{k=1}^N x_k^{-1}\sum_{i=1}^{N}\sum_{j=N+1}^{N+s}\left(\sum_{i=1}^N(1-K_{ij})f+x_j\frac{(1-K_{ij})f}{x_i-x_j}
   \right)|_{x_{N+1}=\ldots= x_{N+s}=0}= \\
   & \omega_{N+1,N+s}^\perp\prod_{k=1}^N x_k^{-1}\sum_{i=1}^{N}\sum_{j=N+1}^{N+s}\left(\sum_{i=1}^N(1-K_{ij})f
   \right)|_{x_{N+1}=\ldots= x_{N+s}=0}=\\
    & \omega_{N+1,N+s}^\perp\prod_{k=1}^N x_k^{-1}\sum_{i=1}^{N}\sum_{j=N+1}^{N+s}\left(\sum_{i=1}^N(1+P_{ij})f
   \right)|_{x_{N+1}=\ldots= x_{N+s}=0}.
  \end{align*} 
  The latter sum equals $N$ times $\omega_N f(x_1,\ldots, x_{N+s})$ due to the definition of $\omega_{N+1,N+s}^\perp$. This  shows that the order of the sum of difference parted of Dunkl operators is also two. Combinations of the above arguments should show that the order of the sums of $m$-th powers of Dunkl operators has the order $m+1$.
  
 \setcounter{equation}{0}
\section{Appendix}
Here we present the expressions for the first densities $\T_{ab,n}(z)$ (\ref{dens}) $n=0,1,2$ for the Yangian generators. There will be given two types of expressions for each density,
the first answer is a normal ordered combination of fermionic fields $\Psi_c(z)$,  $\Psi_d^*(z)$, 
the second is not normal ordered, it is given in terms the affine Lie algebra $\widehat{\gl}_s$ generators.

Now we introduce several notations. Denote by $\T_{ab}^{kl}(z)$ a coefficient of $\beta^{-l}$ in $\T_{ab,k+l}(z)$:
$$
\T_{ab,n}(z)=\sum_{l=0}^n \beta^{-l} \T_{ab}^{n-l,l}(z).
$$
Denote by $E_{ab}(z)$ a generating functions for the elements of the affine Lie algebra $\widehat{\gl}_s$ :
$$
E_{ab}(z)=\sum_n E_{ab,n} z^n=\fnormal z\U{a}^*(z)\U{b}(z)\fnormal.
$$
For a formal series $f(z)=\sum_{n\in\Z}f_nz^n$ we denote by $f(z)_+$ the series $$f(z)_+=\sum_{n\geq 0}f_nz^n=\int\limits_{\substack{u\around 0\\|u|\gg|z|}} du\ \frac{f(u)}{u-z}$$ and by
$f(z)_-$ the series
$$f(z)_-=\sum_{n< 0}f_nz^n=\int\limits_{\substack{u\around 0\\|u|\ll|z|}} du\ \frac{f(u)}{z-u}.$$
For $n=0$ we simply have
\beq\label{tab00}
\T_{ab,0}(z)=\T_{ab}^{0,0}(z)=\fnormal\U{a}^*(z)\U{b}(z)\fnormal=\frac{1}{z}E_{ab}(z).
\eeq
\line(1,0){500}\\
For $n=1 $
$$
\T_{ab,1}(z)=\beta^{-1}\T_{ab}^{0,1}(z)+\T_{ab}^{1,0}(z).
$$
We distinguish the answers for diagonal $\T_{aa,n}(z)$ and nondiagonal part  $\T_{ab,n}(z)$ , where $a\neq b$.  Firstly we present the expressions for nondiagonal elements $a\neq b$ as normal ordered combination of fermionic fields:
$$
\T_{ab}^{0,1}(z)=\fnormal\Psi^*_a(z)z\frac{\partial}{\partial z}\Psi_b(z)\fnormal,
$$
\begin{align*}
\T_{ab}^{1,0}(z)= &\fnormal\sum_{c=1}^s z\Psi^*_a(z)\Psi_b(z)\left( \Psi^*_c(z)\Psi_c(z)\right)_-+\sum_{c=1}^s z\Psi^*_a(z)\Psi_c(z)\left( \Psi^*_c(z)\Psi_b(z)\right)_-\\
 & +(s+1)\ \Psi^*_a(z)\left(z\frac{\partial}{\partial z}\Psi_b(z)\right)_--\Psi_b(z)\left(z\frac{\partial}{\partial z} \Psi_a^*(z)\right)_+\fnormal.
\end{align*}
The bosonic answer has the recurrent form, we express it from $\T_{ab}^{0,0}(z)$ (\ref{tab00}):
$$
\T_{ab}^{0,1}(z)=\int\limits_{w\around z}\frac{dw}{(w-z)} \ E_{aa}(z)\T_{ab}^{0,0}(w),
$$
\begin{align*}
\T_{ab}^{1,0}(z)= &\sum_{c=1}^s \int\limits_{\substack{w\around 0\\|w|\ll|z|}}\frac{dw}{(z-w)}E_{ac}(z)\T_{cb}^{0,0}(w)+\sum_{c=1}^s\int\limits_{\substack{w\around 0\\|w|\ll|z|}}\frac{zdw}{w(z-w)}\T_{ab}^{0,0}(z)E_{cc}(w)-\\
&-\int\limits_{w\around z}\frac{zdw}{w(w-z)}\T_{ab}^{0,0}(z)E_{aa}(w).
\end{align*}
\line(1,0){500}\\
For diagonal elements in case $n=1$ we present the answers in the same way, firstly as a normal ordered combination of fermionic fields:
$$
\T_{aa}^{0,1}(z)=\fnormal\Psi^*_a(z)z\frac{\partial}{\partial z}\Psi_a(z)\fnormal,
$$
\begin{align*}
\T_{aa}^{1,0}(z)= &\fnormal\sum_{b=1}^s z\Psi^*_a(z)\Psi_a(z)\left( \Psi^*_b(z)\Psi_b(z)\right)_-+\sum_{b=1}^s z\Psi^*_a(z)\Psi_b(z)\left( \Psi^*_b(z)\Psi_a(z)\right)_-\\
 & -\sum_{b=1}^s\Psi_b(z)\left(z\frac{\partial}{\partial z} \Psi_b^*(z)\right)_++(s+1)\ \Psi^*_a(z)\left(z\frac{\partial}{\partial z}\Psi_a(z)\right)_--\Psi_a(z)\left(z\frac{\partial}{\partial z} \Psi_a^*(z)\right)_+\fnormal.
\end{align*}
Then the recurrent answer from previous densities  in terms the affine Lie algebra $\widehat{\gl}_s$ generators:
$$
\T_{aa}^{0,1}(z)=\frac{1}{2}\int\limits_{w\around z}\frac{dw}{(w-z)} \ E_{aa}(z)\T_{aa}^{0,0}(w)+\frac{1}{2}\int \limits_{w\around z}\frac{zdw}{(w-z)^2}\T_{aa}^{0,0}(w),
$$
\begin{align*}
\T_{aa}^{1,0}(z)= &\sum_{c=1}^s \int\limits_{\substack{w\around 0\\|w|\ll|z|}}\frac{dw}{(z-w)}E_{ac}(z)\T_{ca}^{0,0}(w)+\sum_{c=1}^s\int\limits_{\substack{w\around 0\\|w|\ll|z|}}\frac{zdw}{w(z-w)}\T_{aa}^{0,0}(z)E_{cc}(w)-\\
&-\sum_{c=1}^s\left(\int\limits_{w\around z}\frac{dw}{(w-z)} \ E_{cc}(z)\T_{cc}^{0,0}(w)-\T_{cc}^{0,1}(z)\right)+\int\limits_{w\around z}\frac{zdw}{(w-z)^2}\T_{aa}^{0,0}(w)-T_{aa}^{0,1}(z).
\end{align*}
\line(1,0){500}\\
For $n=2$
$$
\T_{ab,2}(z)=\beta^{-2}\T_{ab}^{0,2}(z)+\beta^{-1}\T_{ab}^{1,1}(z)+\T_{ab}^{2,0}(z)
$$
We split $\T_{ab}^{1,1}(z)$ into two summands :
$$
\T_{ab}^{1,1}(z)=(\T_{ab}^{1,1})'(z)+(\T_{ab}^{1,1})''(z).
$$
Here $(\T_{ab}^{1,1})'(z)$ means that firstly we apply $z\frac{\partial}{\partial z}$ and then the difference part of the Dunkl operator, $(\T_{ab}^{1,1})''(z)$ backwards.
$$
\T_{ab}^{0,2}(z)=\fnormal\Psi^*_a(z)\left(z\frac{\partial}{\partial z}\right)^2\Psi_b(z)\fnormal
$$
\begin{align*}
&(\T_{ab}^{1,1})'(z)= \fnormal\sum_{c=1}^s z\Psi^*_a(z)\left(z\frac{\partial}{\partial z} \Psi_b(z)\right)\left( \Psi^*_c(z)\Psi_c(z)\right)_-
 +\sum_{c=1}^{s} z \Psi^*_a(z) \Psi_c(z)\left( \Psi^*_c(z)z\frac{\partial}{\partial z} \Psi_b(z)\right)_- \\
 &+\left(s+\frac{1}{2}\right)\ \Psi^*_a(z)\left(\left(z\frac{\partial}{\partial z}\right)^2\Psi_b(z)\right)_- -\left(z\frac{\partial}{\partial z}\Psi_b(z)\right)\left(z\frac{\partial}{\partial z} \Psi_a^*(z)\right)_+-\frac{1}{2} \Psi^*_a(z)\left(z\frac{\partial}{\partial z}\Psi_b(z)\right)_-\fnormal
 \end{align*}
\begin{align*}
(\T_{ab}^{1,1})''(z)= &\fnormal\sum_{c=1}^s \Psi^*_a(z)z\frac{\partial}{\partial z}\left(z \Psi_c(z)\left( \Psi^*_c(z)\Psi_b(z)\right)_-\right)
 +\sum_{c=1}^s \Psi^*_a(z)z\frac{\partial}{\partial z}\left(z \Psi_b(z)\left( \Psi^*_c(z)\Psi_c(z)\right)_-\right) \\
 &+(s+1)\ \Psi^*_a(z)\left(\left(z\frac{\partial}{\partial z}\right)^2\Psi_b(z)\right)_- -\frac{1}{2}\Psi_b(z)\left(z\frac{\partial}{\partial z} \Psi_a^*(z)\right)_+\\
 &-\left(z\frac{\partial}{\partial z}\Psi_b(z)\right)\left(z\frac{\partial}{\partial z} \Psi_a^*(z)\right)_+-\frac{1}{2}\Psi_b(z)\left(\left(z\frac{\partial}{\partial z}\right) ^2\Psi_a^*(z)\right)_+\fnormal
 \end{align*}
The recurrent formula from previous densities in terms the affine Lie algebra $\widehat{\gl}_s$ generators:
 $$
\T_{ab}^{0,2}(z)=\int\limits_{w\around z}\frac{dw}{(w-z)} \ E_{aa}(z)\T_{ab}^{0,1}(w)
$$
\begin{align*}
 (\T_{ab}^{1,1})'(z)= &\sum_{c=1}^s \int\limits_{\substack{w\around 0\\|w|\ll|z|}}\frac{dw}{(z-w)}E_{ac}(z)\T_{cb}^{0,1}(w)+\sum_{c=1}^s\int\limits_{\substack{w\around 0\\|w|\ll|z|}}\frac{zdw}{w(z-w)}\T_{ab}^{0,1}(z)E_{cc}(w)-\\
&-\int\limits_{w\around z}\frac{zdw}{w(w-z)}\T_{ab}^{0,1}(z)E_{aa}(w)
\end{align*}
\begin{align*}
(\T_{ab}^{1,1})''(z)= &\int\limits_{w\around z}\frac{dw}{(w-z)}E_{aa}(z)\T_{ab}^{1,0}(w)+\int\limits_{\substack{w\around 0\\|w|\ll|z|}}\frac{zdw}{w(z-w)}\left(\T_{aa}^{0,0}(z)+z\frac{\partial}{\partial z}
 \T_{aa}^{0,0}(z)-\T_{aa}^{0,1}(z)\right)E_{ab}(w)-\\
&-\int\limits_{\substack{w\around 0\\|w|\ll|z|}}\frac{zdw}{w(z-w)^2}E_{aa}(z)E_{ab}(w)-\int\limits_{w\around z}\frac{zdw}{w(w-z)}\T_{ab}^{0,1}(z)E_{aa}(w)\\
&+\int \limits_{w\around z}\frac{zdw}{(w-z)^3}E_{ab}(w)-\int\limits_{\substack{w\around 0\\|w|\ll|z|}}\frac{zdw}{(z-w)^3}E_{ab}(w)-\frac{1}{2} \T_{ab}^{0,2}(z)-\frac{1}{2} \T_{ab}^{0,1}(z)
\end{align*}
 \line(1,0){500}\\
For diagonal elements in case $n=2$ we have more complicated formulas:
$$
\T_{aa}^{0,2}(z)=\fnormal\Psi^*_a(z)\left(z\frac{\partial}{\partial z}\right)^2\Psi_a(z)\fnormal
$$
\begin{align*}
(\T_{aa}^{1,1})'(z)= &\fnormal\sum_{c=1}^s z\Psi^*_a(z)\left(z\frac{\partial}{\partial z} \Psi_a(z)\right)\left( \Psi^*_c(z)\Psi_c(z)\right)_-
 +\sum_{c=1}^{s} z \Psi^*_a(z) \Psi_c(z)\left( \Psi^*_c(z)z\frac{\partial}{\partial z} \Psi_a(z)\right)_- \\
 &+\left(s+\frac{1}{2}\right)\ \Psi^*_a(z)\left(\left(z\frac{\partial}{\partial z}\right)^2\Psi_a(z)\right)_- +\frac{1}{2}\sum_{c=1}^s\Psi_c(z)\left(\left(z\frac{\partial}{\partial z}\right)^2 \Psi_c^*(z)\right)_+\\ 
 &+\frac{1}{2}\sum_{c=1}^s\Psi_c(z)\left(z\frac{\partial}{\partial z} \Psi_c^*(z)\right)_+-\left(z\frac{\partial}{\partial z}\Psi_a(z)\right)\left(z\frac{\partial}{\partial z} \Psi_a^*(z)\right)_+\fnormal
 \end{align*}
\begin{align*}
(\T_{aa}^{1,1})''(z)= &\fnormal\sum_{b=1}^s \Psi^*_a(z)z\frac{\partial}{\partial z}\left(z \Psi_a(z)\left( \Psi^*_b(z)\Psi_b(z)\right)_-\right)
 +\sum_{b=1}^s \Psi^*_a(z)z\frac{\partial}{\partial z}\left(z \Psi_b(z)\left( \Psi^*_b(z)\Psi_a(z)\right)_-\right) \\
 & -\frac{1}{2}\sum_{b=1}^s\Psi_b(z)\left(z\frac{\partial}{\partial z} \Psi_b^*(z)\right)_+-\frac{1}{2}\sum_{b=1}^s\Psi_b(z)\left(\left(z\frac{\partial}{\partial z}\right) ^2\Psi_b^*(z)\right)_+\\
 &-\sum_{b=1}^s\left(z\frac{\partial}{\partial z}\Psi_b(z)\right)\left(z\frac{\partial}{\partial z} \Psi_b^*(z)\right)_++(s+1)\ \Psi^*_a(z)\left(\left(z\frac{\partial}{\partial z}\right)^2\Psi_a(z)\right)_-\\
 & -\frac{1}{2}\Psi_a(z)\left(z\frac{\partial}{\partial z} \Psi_a^*(z)\right)_+-\frac{1}{2}\Psi_a(z)\left(\left(z\frac{\partial}{\partial z}\right) ^2\Psi_a^*(z)\right)_+-\left(z\frac{\partial}{\partial z}\Psi_a(z)\right)\left(z\frac{\partial}{\partial z} \Psi_a^*(z)\right)_+\fnormal\\
 \end{align*}
$$
\T_{aa}^{0,2}(z)=\frac{2}{3}\int\limits_{w\around z}\frac{dw}{(w-z)} \ E_{aa}(z)\T_{aa}^{0,1}(w)-\frac{2}{3}\int\limits_{w\around z}\frac{zdw}{(w-z)^3} E_{aa}(w)+\frac{2}{3}\int\limits_{w\around z}\frac{zdw}{(w-z)^2} \T_{aa}^{0,1}(w)
+\frac{1}{3}\T_{aa}^{0,1}(z)
$$
\begin{align*}
(\T_{aa}^{1,1})'(z)= &\sum_{c=1}^s \int\limits_{\substack{w\around 0\\|w|\ll|z|}}\frac{dw}{(z-w)}E_{ac}(z)\T_{ca}^{0,1}(w)+\sum_{c=1}^s\int\limits_{\substack{w\around 0\\|w|\ll|z|}}\frac{zdw}{w(z-w)}\T_{aa}^{0,1}(z)E_{cc}(w)-\\
&-\sum_{c=1}^s\left(\int\limits_{w\around z}\frac{dw}{(w-z)} \ E_{cc}(z)\T_{cc}^{0,1}(w)-\T_{cc}^{0,2}(z)\right)+\int\limits_{w\around z}\frac{zdw}{(w-z)^2}\T_{aa}^{0,1}(w)-T_{aa}^{0,2}(z)
\end{align*}
\begin{align*}
(\T_{aa}^{1,1})''(z)&= \int\limits_{w\around z}\frac{dw}{(w-z)}E_{aa}(z)\T_{aa}^{1,0}(w)+\\
&+\sum_{b=1}^s\int\limits_{\substack{w\around 0\\|w|\ll|z|}}\frac{zdw}{w(z-w)}\left(\T_{aa}^{0,0}(z)+z\frac{\partial}{\partial z}
 \T_{aa}^{0,0}(z)-\T_{aa}^{0,1}(z)\right)E_{bb}(w)-\\
 &-\sum_{b=1}^s\int\limits_{\substack{w\around 0\\|w|\ll|z|}}\frac{zdw}{w(z-w)^2}E_{aa}(z)E_{bb}(w)-\sum_{b=1}^s\int\limits_{\substack{w\around 0\\|w|\ll|z|}}\frac{zdw}{(z-w)^3}E_{bb}(w)\\
 &+\int\limits_{\substack{w\around 0\\|w|\ll|z|}}\frac{zdw}{w(z-w)}\left(\T_{aa}^{0,0}(z)+z\frac{\partial}{\partial z}
 \T_{aa}^{0,0}(z)-\T_{aa}^{0,1}(z)\right)E_{aa}(w)\\
 &-\int\limits_{\substack{w\around 0\\|w|\ll|z|}}\frac{zdw}{w(z-w)^2}E_{aa}(z)E_{aa}(w)-\int\limits_{\substack{w\around 0\\|w|\ll|z|}}\frac{zdw}{(z-w)^3}E_{aa}(w)\\
&+\sum_{b=1}^s\int\limits_{w\around z}\frac{dw}{(w-z)}\left(\T_{aa}^{0,1}(z)-z\frac{\partial}{\partial z} \T_{aa}^{0,0}(z)\right)E_{bb}(w)\\
&+\sum_{b=1}^s \int \limits_{w\around z}\frac{zdw}{(w-z)^3}E_{bb}(w)-\sum_{b=1}^s\int\limits_{w\around z}\frac{dw}{(w-z)^2}E_{aa}(z)E_{bb}(w)\\
&+s\left(\int\limits_{w\around z}\frac{zdw}{(w-z)^2} \T_{aa}^{0,1}(w)-\int\limits_{w\around z}\frac{zdw}{(w-z)^3} E_{aa}(w)\right)-\frac{s}{2}\T_{aa}^{0,2}(z)+\frac{s}{2}\T_{aa}^{0,1}(z)\\
&+2\int\limits_{w\around z}\frac{zdw}{(w-z)^3} E_{aa}(w)-2\int\limits_{w\around z}\frac{zdw}{(w-z)^2} \T_{aa}^{0,1}(w)+\int\limits_{w\around z}\frac{zdw}{w(w-z)^2} E_{aa}(w)+\T_{aa}^{0,2}(z)
\end{align*}
The density $\T_{11}^{2,0}(z)$ has a cumbersome form and we do not present it here. In scalar case ($s=1$) the matrix coefficient $T_{11}^{2,0}$ is a polynomial in zero mode of the scalar bosonic field:
$$
T_{11}^{2,0}=\frac{1}{6}\left(2 a_0^3-3 a_0^2+a_0\right).
$$
In scalar case the same is for higher orders: the matrix coefficient $T_{11}^{n,0}$ is a polynomial of degree $(n+1)$ in zero mode of the scalar bosonic field \cite{Mat}.
\section*{Acknowledgements} 
The authors are grateful to M.L. Nazarov and E.K.Sklyanin for fruitful
discussions on the subject of the paper. The research by M.M. was carried out within the HSE University Basic Research Program
                                                     and funded jointly by the Russian Academic Excellence Project '5-100'. It was also supported in part by the Simons Foundation.
                                                     S.K appreciates the support of Russian Science Foundation grant, project 16-11-10316 used for the proofs of Propositions 4.1, 4.2, 4.3 section 4.

\end{document}